\documentclass[9pt]{article}
\usepackage{graphicx}

\renewcommand{\in}{\raise -3pt\hbox{\scriptsize in}}
\newcommand{\out}{\raise -3pt\hbox{\scriptsize out}}

\begin{document}

\begin{flushright}
CPHT--RR 064.1104 \\
LPT--04.129
\end{flushright}

\vspace{\baselineskip}

\begin{center}
\textbf{\LARGE On BLM scale fixing in exclusive processes } \\

\vspace{1\baselineskip}
{\large I.V.~Anikin$^{a,b,d}$,\, B.~Pire$^b$,\,
 L.~Szymanowski$^{c,e}$,\, O.V.~Teryaev$^{a}$,\, S.~Wallon$^d$
}
\\
\vspace{1\baselineskip}
{\it ${}^a$\,Bogoliubov Laboratory of Theoretical Physics, JINR, 141980 Dubna,
Russia \\
${}^b$\,CPHT \footnote{Unit{\'e} mixte 7644 du CNRS}, {\'E}cole
Polytechnique, 91128 Palaiseau, France \\
${}^c$\,Soltan Institute for Nuclear Studies, Warsaw, Poland \\
${}^d$\,LPT \footnote{Unit{\'e} mixte 8627 du CNRS}, Universit{\'e} Paris-Sud, 91405-Orsay,
France
 \\
${}^e$\,Phys. Th\'eor. Fondam., Inst. de Physique,
Univ. de Li{\`e}ge, B-4000 Li{\`e}ge, Belgium
\\
}
\vspace{1\baselineskip}
\textbf{Abstract}\\
\vspace{1\baselineskip}
\parbox{0.9\textwidth}
{We discuss the BLM scale fixing procedure in exclusive electroproduction processes in the Bjorken 
regime with rather large $x_B$.
We show that in the case of vector meson production dominated in this case by quark exchange
the usual way to apply the BLM method fails due to singularities present in 
equations fixing the BLM scale.
We argue that the BLM scale should be extracted
from the squared amplitudes which are directly related to observables.}
\end{center}

\vspace{.5cm}
\section*{Introduction}
\vspace{.5cm}

\noindent
The investigation of the quark-gluon dynamics through perturbative calculations is 
most useful to extract from experimentally measurable observables  quantities such as 
parton distributions, generalized parton distributions (GPDs) and (generalized) 
distribution amplitudes (GDAs, DAs). Factorization theorems  allow to calculate
scattering amplitudes in a perturbative way, provided a renormalization scale and a factorization
scale are choosen. Whereas  the observables in 
the extensively studied inclusive reactions
are in general related to an amplitude (as the case of inclusive DIS expressed as the imaginary 
part of the forward virtual Compton scattering reaction), 
exclusive cross sections which have been 
much studied recently are based on a factorization theorem at the amplitude level, 
and thus require to square the
amplitude given by its perturbative expansion. 
Renormalization scale fixing has been the subject of intense studies and different strategies
\cite{opt, BLM} have been put forward to maximize the predictivity of theoretical 
studies  through ensuring the smallness of  corrections related to  higher orders terms
in the perturbative series.  The phenomenological success of these proposals is quite impressive
in a number of 
cases, mostly related to inclusive cross sections or jet physics. With the advent of next to leading
order results, it has been advocated to use these procedures also in hard exclusive 
processes \cite{Belitsky:2001nq, IS}.
Even at the Born level, the hard meson  electroproduction amplitude contains  
$\alpha_S$. Therefore the choice of the renormalization scale $\mu_R$ is very crucial for the 
practical estimations of the observables related to meson electroproduction.  
The first study of a hard electroproduction amplitude including the analysis of the 
next-to-leading orders (NLO) has been implemented by Belitsky and Muller 
\cite{Belitsky:2001nq} for  $\pi^+$ production, {\it i.e.} for $\gamma^* p\to\pi^+ n$ process. 
This study used an appropriate continuation of the NLO calculations known for the electromagnetic pion 
form factor \cite{ffNLO} onto the case of the meson electroproduction process. 

\noindent
In this work,  focusing on the Brodsky-Lepage-Mackenzie (BLM) procedure \cite{BLM}
\footnote{We expect our remarks to be quite general, so that they should apply also to other optimization 
procedures}, 
we examine in detail the consequences of the fact that exclusive processes 
considered in the regime where the quark GPDs are dominant  are 
factorized at the amplitude level
and that  the meson electroproduction amplitude is a complex function. 
Consequently, we are forced to apply the BLM procedure to the real and to the imaginary part
of the scattering amplitude separately, which in general leads to two different
scales. 
Moreover, we show that such a way of  scale fixing,
as has been done in \cite{Belitsky:2001nq} for the $\pi-$meson production, 
leads to unphysical results in case of vector meson production. We propose
a way to modify the BLM procedure in order to avoid such difficulties.

\vspace{.5cm}
\section*{Basics of the BLM procedure}
\vspace{.5cm}

\noindent
The QCD factorization theorem \cite{CFS} states that the amplitude of 
hard meson electroproductions
can be written as
\begin{eqnarray}
\label{AmF}
{\cal A}=\int\limits_{0}^{1} dz \int\limits_{-1}^{1} dx\,
\Phi_M(z,\mu^2_F) \, H(x,z,Q^2,\mu^2_F,\mu^2_R) \,
F(x,\mu^2_F)\equiv
\Phi_M\, \otimes \,H \, \otimes\, F,
\end{eqnarray}
where the parameters $\mu_F^2$ and $\mu^2_R$ are the factorization and
renormalization scales, respectively. 
The scales $\mu_R^2$ and $\mu_F^2$ are in principle independent but often  
it is argued that they can coincide, $\mu_R^2 = \mu_F^2$. 
The arguments in favour of such assumption are 
discussed in, e.g. \cite{RadE}, and we adopt this also in the 
present paper (to simplify notation we omit below subscripts $R$ and $F$).
In Eq.(\ref{AmF}), $H$ is the hard part of amplitude which is controlled by  perturbative QCD.
The meson distribution amplitude $\Phi_M$ describes the transition from  the partons
to the meson, and $F$ denotes the GPDs which are related to
nonperturbative matrix elements of bilocal operators between different hadronic states.

\noindent
The product $\Phi_H\, \otimes \,H \, \otimes\, F$ in (\ref{AmF})
is, generally speaking, independent of the
particular choice of the parameter $\mu^2$. However, this independence is broken 
once we limit ourselves to the first few terms in an expansion over the coupling
constant $\alpha_S$. In this case, the theoretical ambiguity of the choice of the parameter $\mu^2$
 emerges. The goal is to choose the parameter $\mu^2$  such as to ensure that pretty 
small contributions will arise from the next order  corrections.  
Out of the several possible ways  to hope to reach that goal, the BLM 
procedure \cite{BLM} begins with separating out  the terms which are proportional to the 
one-loop $\beta-$function, $\beta_0=11-2/3 N_F$, appearing  in the NLO terms. 
The amplitude (\ref{AmF}) including the NLO corrections with separated terms proportional to $\beta_0$
 reads
\begin{eqnarray}
\label{AmF2}
{\cal A}=\alpha_S(\mu^2)\,{\cal A}^{{\rm LO}}(Q^2) +
\alpha_S^2(\mu^2)\, \frac{\beta_0}{4\pi} \Biggl\{
\biggl[ C - {\rm ln}\frac{Q^2}{\mu^2} \biggr] \,{\cal A}^{{\rm LO}}(Q^2) +
\tilde{\cal A}^{{\rm NLO},(\beta)}(Q^2)\Biggr\} + ... ,
\end{eqnarray}
where the ellipsis stand for the terms of the NLO corrections which do not explicitly contain
 $\beta_0$.  
In (\ref{AmF2}), the value of the constant $C$ depends on the kind of produced mesons. 
As pointed out in \cite{Belitsky:2001nq}, \cite{ffNLO}
the exact expressions, in the quark sector,
with the NLO corrections may be obtained by a suitable substitution from 
the well-known results for the pion electromagnetic form factor.

\noindent
Due to the renormalization group equations the coupling constant takes the form
\begin{eqnarray}
\label{RG}
\alpha_S(\mu^2)=\frac{1}{(\beta_0/4\pi)\,{\rm ln}(\mu^2/\Lambda^2_{QCD})}=
\frac{\alpha_S(\mu^2_0)}{1+(\beta_0/4\pi)\,\alpha_S(\mu^2_0)\,
{\rm ln}(\mu^2/\mu^2_0)}.
\end{eqnarray}
We insert this expression into the amplitude (\ref{AmF2})
and then expand it in powers of $\alpha_S(\mu^2_0)$.
Retaining 
 the terms which are proportional to  $\beta_0$, we get
\begin{eqnarray}
\label{ExpAmp}
&&{\cal A}=\alpha_S(\mu^2_0)\,{\cal A}^{{\rm LO}}(Q^2) -
\alpha_S^2(\mu^2_0)\,\frac{\beta_0}{4\pi}\,{\rm ln}\frac{\mu^2}{\mu^2_0}
\,{\cal A}^{{\rm LO}}(Q^2) +
\nonumber\\
&&\alpha_S^2(\mu^2_0)\,\frac{\beta_0}{4\pi} \Biggl\{
\biggl[ C - {\rm ln}\frac{Q^2}{\mu^2} \biggr] \,{\cal A}^{{\rm LO}}(Q^2) +
\tilde{\cal A}^{{\rm NLO},(\beta)}(Q^2)\Biggr\} + ...
\end{eqnarray} 

\noindent
The BLM procedure consists in the choice of such $\mu_0^2$ for which the whole term 
proportional to $\beta_0$ in Eq.~(\ref{ExpAmp}) vanishes, i.e. 
\begin{eqnarray}
\alpha_S^2(\mu^2_0)\,\frac{\beta_0}{4\pi}\,{\rm ln}\frac{\mu^2}{\mu^2_0}
\,{\cal A}^{{\rm LO}} -
\alpha_S^2(\mu^2_0)\, \frac{\beta_0}{4\pi} \Biggl\{
\biggl[ C - {\rm ln}\frac{Q^2}{\mu^2} \biggr] \,{\cal A}^{{\rm LO}} +
\tilde{\cal A}^{{\rm NLO},(\beta)}\Biggr\}=0,
\end{eqnarray}
from which it follows that the BLM scale $\mu^2_{BLM}$ is equal to  
\begin{eqnarray}
\mu^2_0=\mu^2_{BLM}=Q^2\, e^{-f}, \quad 
f= C+\frac{\tilde{\cal A}^{{\rm NLO},(\beta)}}{{\cal A}^{{\rm LO}}}.
\end{eqnarray}

\noindent
If the scattering amplitude is real, as in the case of the spacelike pion form factor, the above
procedure leads just to the one $\mu^2_{BLM}$ scale. But already in the $\pi^+-$meson electroproduction
the scattering amplitude is a complex function and if one applies the BLM procedure separately
both for real and imaginary parts
\footnote{The exchange of the on-shell (light-like) gluon 
is entirely responsible for the imaginary part of the amplitude.
This, however, does not break the factorization.}, this results in two different scales. 
The situation starts to be
even worse in the case of the vector meson electroproduction which we discuss in the next section.

\vspace{.5cm}
\section*{Extraction of the BLM scale from the amplitudes}
\vspace{.5cm}

\noindent
We now focus on the numerical estimation of the renormalization scales extracted 
them from the amplitudes of the vector mesons electroproduction. 
We consider only spin non-flip quark GPD $H$ of the nucleon target and neglect
spin-flip GPD $E$. The reason is that: a) $E$ function gives very small 
contribution to the
scattering amplitude, b) its form is very model dependent.

\noindent
Consider the NLO terms of amplitude (\ref{AmF2}) containing the $\beta_0$ coefficient
\begin{eqnarray}
\label{blm-1}
&&\biggl[ C - {\rm ln}\frac{Q^2}{\mu^2} \biggr] \,{\cal A}^{{\rm LO}}(Q^2) +
\tilde{\cal A}^{{\rm NLO},(\beta)}(Q^2)\sim
\int\limits_{0}^{1}dz\, \int\limits_{-1}^{1} dx\,
\left\{ \Phi_\rho(z)\atop \Phi_H(z)\right\}
H^p (x,\xi,t_{min})
\\
&&\,\biggl\{\,
\frac{2\xi}{z(\xi+x)}\biggl[
\frac{5}{3}- {\rm ln} (\frac{z(\xi+x)}{2\xi}) - {\rm ln}(\frac{Q^2}{\mu^2})
\biggr]\mp (z\to \bar z;\, x\to -x)
\,\biggr\} =0.
\nonumber
\end{eqnarray}
In Eqn. (\ref{blm-1}), the distribution amplitudes $\Phi_\rho(z)$ and $\Phi_H(z)$
correspond to the $\rho$ and hybrid mesons, respectively.
The hybrid meson is a charge conjugation even state ($J^{PC}=1^{-+}$), as studied in 
\cite{APSTW1}.
The function $H^p (x,\xi,t_{min})$ stand for the corresponding GPD's and is defined as
\begin{eqnarray}
H^{p}(x,\xi,t)=\frac{1}{\sqrt{2}}\biggl( 
e_u H^{u}(x,\xi,t) - e_d H^{d}(x,\xi,t)
\biggr).
\end{eqnarray}
The meson distribution amplitudes $\Phi_\rho$ and $\Phi_H$ may be understood as the asymptotic 
functions which are (see, for instance \cite{APSTW1}):
\begin{eqnarray}
\label{asfs}
\Phi_\rho(z)=6z\bar z, \quad \Phi_H(z)=30z\bar z(1-2z).
\end{eqnarray}
Note that the NLO evolution effects for the meson distribution amplitudes seem to be small
and we omit the consideration of such effects.  

\noindent
The GPD's, in (\ref{blm-1}), can be modelled using the Radyushkin model \cite{Rad} which
ensures the agreement with the
forward limit and the corresponding sum rules for the moments.
According to this ansatz, the GPD's are expressed with the help of double distributions 
$F^q(x,y;t)$:
\begin{eqnarray}
\label{GPD}
H^q(x,\xi,t)=\frac{\theta(\xi+x)}{1+\xi}
\int\limits_{0}^{{\rm min}\{ \frac{x+\xi}{2\xi},
\frac{1-x}{1-\xi} \}} dy\,
F^q(x_+, y, t) -
\frac{\theta(\xi-x)}{1+\xi}
\int\limits_{0}^{{\rm min}\{ \frac{\xi-x}{2\xi},
\frac{1+x}{1-\xi} \}} dy\,
F^{\bar q}(x_-, y, t),
\end{eqnarray}
where 
\begin{eqnarray}
x_+=\frac{x+\xi-2\xi y}{1+\xi}, \quad x_-=\frac{\xi-x-2\xi y}{1+\xi}.
\end{eqnarray}
For the double distribution $F^q(X,Y;t)$, we assume the ansatz suggested by Radyushkin 
\cite{Rad}:
\begin{eqnarray}
\label{DD}
F^q(X,Y;t)=\frac{F_1^q(t)}{F_1^q(0)}\, q(X)\, 6\frac{Y(1-X-Y)}{(1-X)^3},
\end{eqnarray}
where the forward (anti)quark distribution is taken from the parameretization of \cite{MRST98}.
Note that, in (\ref{DD}), $t_{min}$ is different from zero and is equal to $-4m^2_N\xi^2/(1-\xi^2)$
(see, for instance, \cite{APSTW2}).
A similar expression gives the anti-quark contribution.

\noindent
As shown in \cite{PW-LD},
the definition of the double distribution is not completely compatible with
the structure of the corresponding matrix elements; introducing D-terms
restores the self-consistency of this representation. Taking into account these D-terms
with a factorized $t$-dependence as in Eq. (\ref{DD}),
the GPD's (\ref{GPD}) are modified into :
\begin{eqnarray}
\label{eqnarray}
H^q_D(x,\xi,t)=H^q(x,\xi,t)+\theta(\xi-|x|)\frac{D(x/\xi, t)}{N_f},
\end{eqnarray}
where $D(x/\xi,0)$ is given by ($C_n^{3/2}$ are Gegenbauer polynomials)
\begin{eqnarray}
\label{Dterm}
D(\alpha)=-4(1-\alpha^2)\biggl\{
C_1^{3/2}(\alpha)+0.3 C_3^{3/2}(\alpha)+0.1 C_5^{3/2}(\alpha)
\biggr\}
\end{eqnarray}
with
\begin{eqnarray}
\alpha=\frac{x}{\xi}, \quad D(\alpha)=-D(-\alpha).
\end{eqnarray}
These D-terms do contribute to the interval $-\xi < x < \xi$ of
the GPD's. Besides, due to the anti-symmetric properties of (\ref{Dterm}),
the D-terms are important only for the charge conjugation odd vector 
meson ({\it e.g.} $\rho$) production amplitude rather than 
for exotic hybrid meson ($J^{PC}=1^{-+}$) production amplitude \cite{APSTW1}.

\noindent
As aforementioned, the amplitudes of the mesons electroproductions contain 
real and imaginary parts. So, we now come to the estimation of BLM scales for the real 
and imaginary parts of amplitudes, separately. Repeating the procedure pointed out in 
the preceding section, we derive that the BLM scales read
\begin{eqnarray}
\label{muRe}
&&\mu^2_{(Re)}=\, e^{-f_{Re}(\xi)}\, Q^2, 
\\
\label{muIm}
&&\mu^2_{(Im)}=\, e^{-f_{Im}(\xi)}\, Q^2,
\end{eqnarray}
where
\begin{eqnarray}
\label{fre}
&&f_{Re}^\rho (\xi)=\frac{19}{6}-\frac{H^{Re}_2(\xi)}{H^{Re}_1(\xi)},
\\
\label{fim}
&&f_{Im}^\rho (\xi)=\frac{19}{6}-\frac{H^{Im}_2(\xi)}{H^{Im}_1(\xi)}.
\end{eqnarray}
for the $\rho$ meson production, and
\begin{eqnarray}
\label{freH}
&&f_{Re}^H(\xi)=\frac{9}{2}-\frac{G^{Re}_2(\xi)}{G^{Re}_1(\xi)},
\\
\label{fimH}
&&f_{Im}^H(\xi)=\frac{9}{2}-\frac{G^{Im}_2(\xi)}{G^{Im}_1(\xi)}.
\end{eqnarray}
for the hybrid meson production.
The explicit expressions for functions $H^{Re}_i(\xi)$,
$H^{Im}_i(\xi)$, $G^{Re}_i(\xi)$ and $G^{Im}_i(\xi)$ in (\ref{fre})--(\ref{fimH})
can be found in the Appendix.
\begin{figure}
$$\rotatebox{270}{\includegraphics[width=8cm]{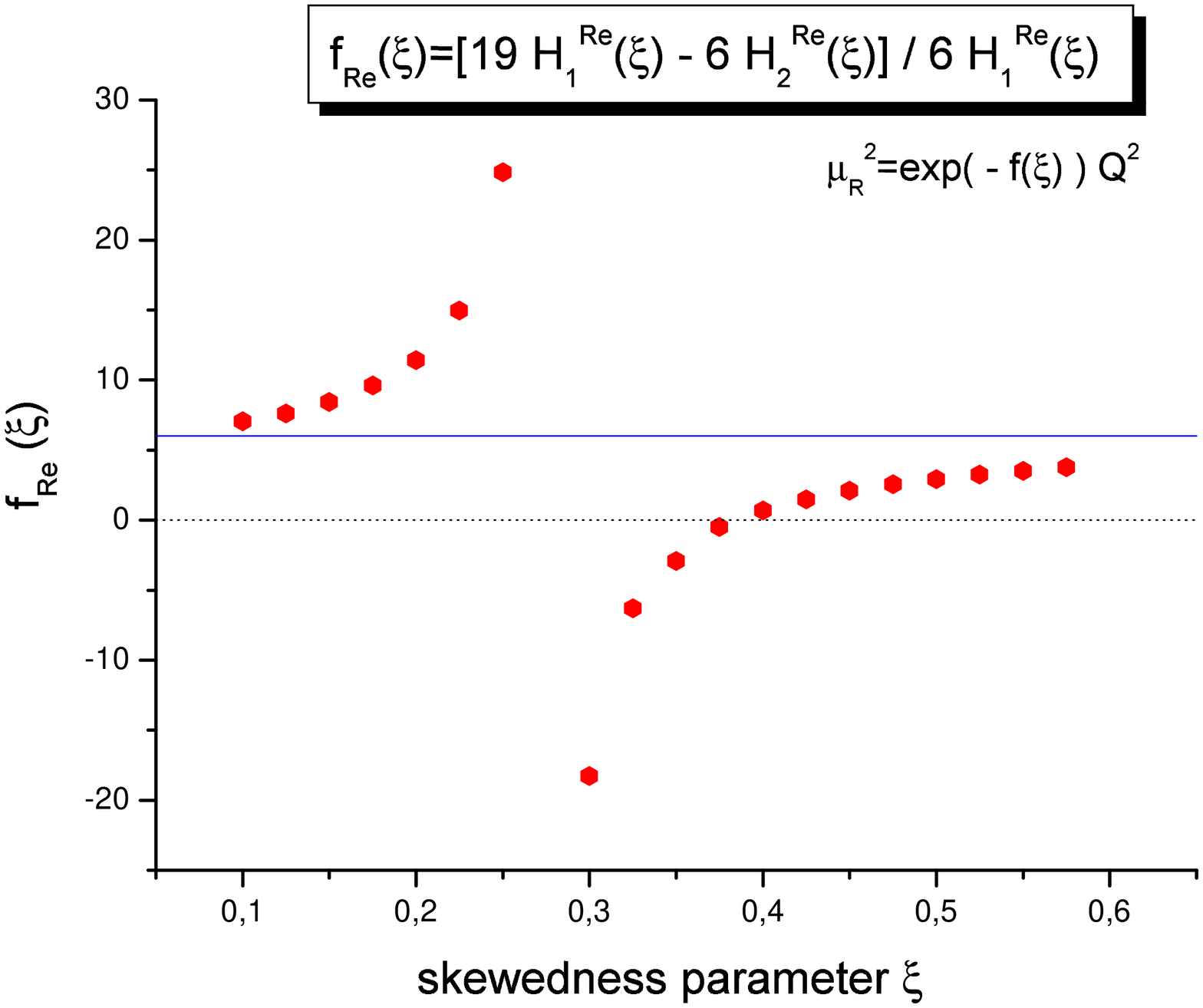}}$$
\caption{The $\rho$ meson production: the BLM scale for the real part of the amplitude.}
\label{blm1}
\end{figure}
\begin{figure}
$$\rotatebox{270}{\includegraphics[width=8cm]{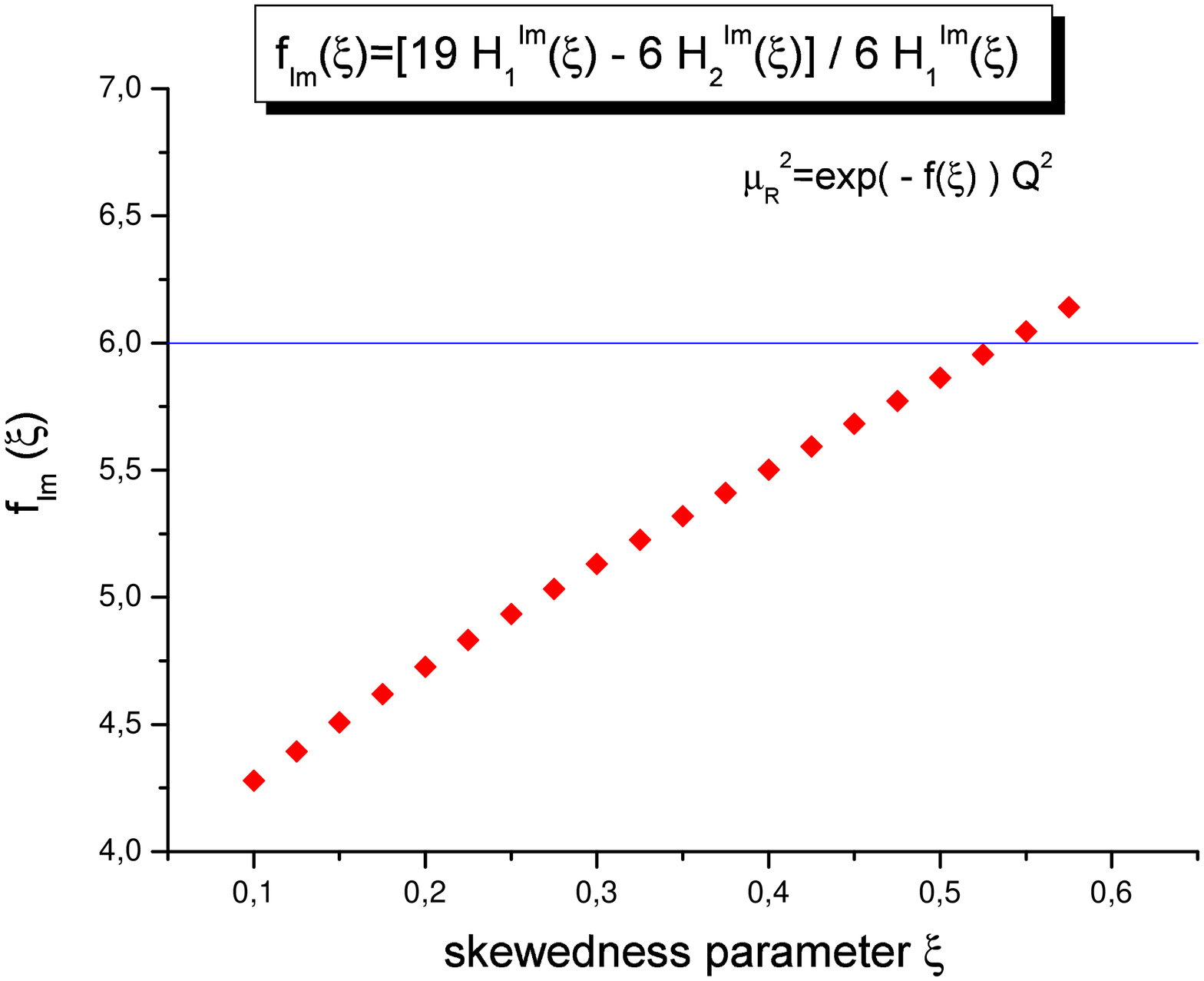}}$$
\caption{The $\rho$ meson production: the BLM scale for the imaginary part of the amplitude.}
\label{blm2}
\end{figure}

\noindent
The investigation of the $\rho$ meson scale with (\ref{fre})
shows that the extraction of the BLM scale from the expression for the amplitude
meet  difficulties. Indeed, as one can see on Fig. \ref{blm1},
the $\rho$ meson function $f_{Re}^\rho(\xi)$ has an unphysical singularity
owing to the fact that the denominator in (\ref{fre}) may vanish
(see the exact expression for $H^{Re}_i(\xi)$ in Appendix).
Indeed, let us dwell on the equation (\ref{Hre1}) from the Appendix which is 
the denominator  of Eq.~(\ref{fre}). The integrand of (\ref{Hre1}) 
is a sign-changing function: 
the integrand is negative in the region $-\xi < x < \xi$
while it is positive in the regions $-1 < x < -\xi$ and $\xi < x < 1$. 
Qualitatively, it is clear that at some value of 
$\xi$ the positive contribution to the whole integral (which is taken in the sense of the Cauchy
pincipal value) will be equilibrated by the negative one. The numerical calculations
show that the dangerous singularity appears for  $\xi\approx 0.27$. 
This value depends on the parameterization of GPDs but the existence of a singularity
is a model-independent result of our analysis.   
Concerning the function $f_{Im}^\rho(\xi)$ in (\ref{fim}), 
as one can see from Fig. \ref{blm2}, this function is always analytical. 

\noindent
For the hybrid meson scale, the situation is analogous. In this case, the sign of the integrand of 
Eq.~(\ref{Hre1H}) of the Appendix is opposite to the one of the integrand in (\ref{Hre1}):
the integrand is positive in the region $-\xi < x < \xi$ and is negative in the regions
$\xi < x < 1$. Thus, we may expect that the whole integral (\ref{Hre1H})
can add up to the zeroth value. The BLM scale corresponding to the imaginary part of the 
hybrid meson production amplitude is again an analytical function.

\noindent
It is instructive to compare the BLM scale fixing for vector meson production with the one for
the $\pi^+$ meson production \cite{Belitsky:2001nq}. In this second case such a singularity does not
appear and equations fixing BLM scale, both for real and imaginary parts of the scattering 
amplitude, are analytical.  
Indeed,  the corresponding integral determining the BLM scale for 
the real part of the $\pi^+$ meson production amplitude, {\it i.e.}
\begin{eqnarray}
\label{HrePI}
{\cal P}\int\limits_{-1}^{1}dx \tilde H^{ud}_{\pi^+}(x,\xi,t_{min})
\Biggl[\frac{e_d}{\xi+x}-\frac{e_u}{\xi-x}
\Biggr],
\end{eqnarray}
will never be equal to zero.

\noindent
Summarizing this section, 
the physical causes for the appearance of the singularity 
are the C-parity conservation and the factorization in hard reactions
and the mathematical evidence is the sign-changing integrands of 
(\ref{Hre1}) and (\ref{Hre1H}).
Moreover, the vanishing of the first order term in the real part only does not imply that the 
scale of the gluon propagator vanishes.
We thus see that the usual BLM scale fixing needs to be
modified in the case of vector meson production. We propose to extract this scale directly 
from an observable, i.e. starting from the square of the scattering amplitude.

\vspace{.5cm}
\section*{Extraction of the BLM scale from the cross section}
\vspace{.5cm}

\begin{figure}
$$\rotatebox{270}{\includegraphics[width=8cm]{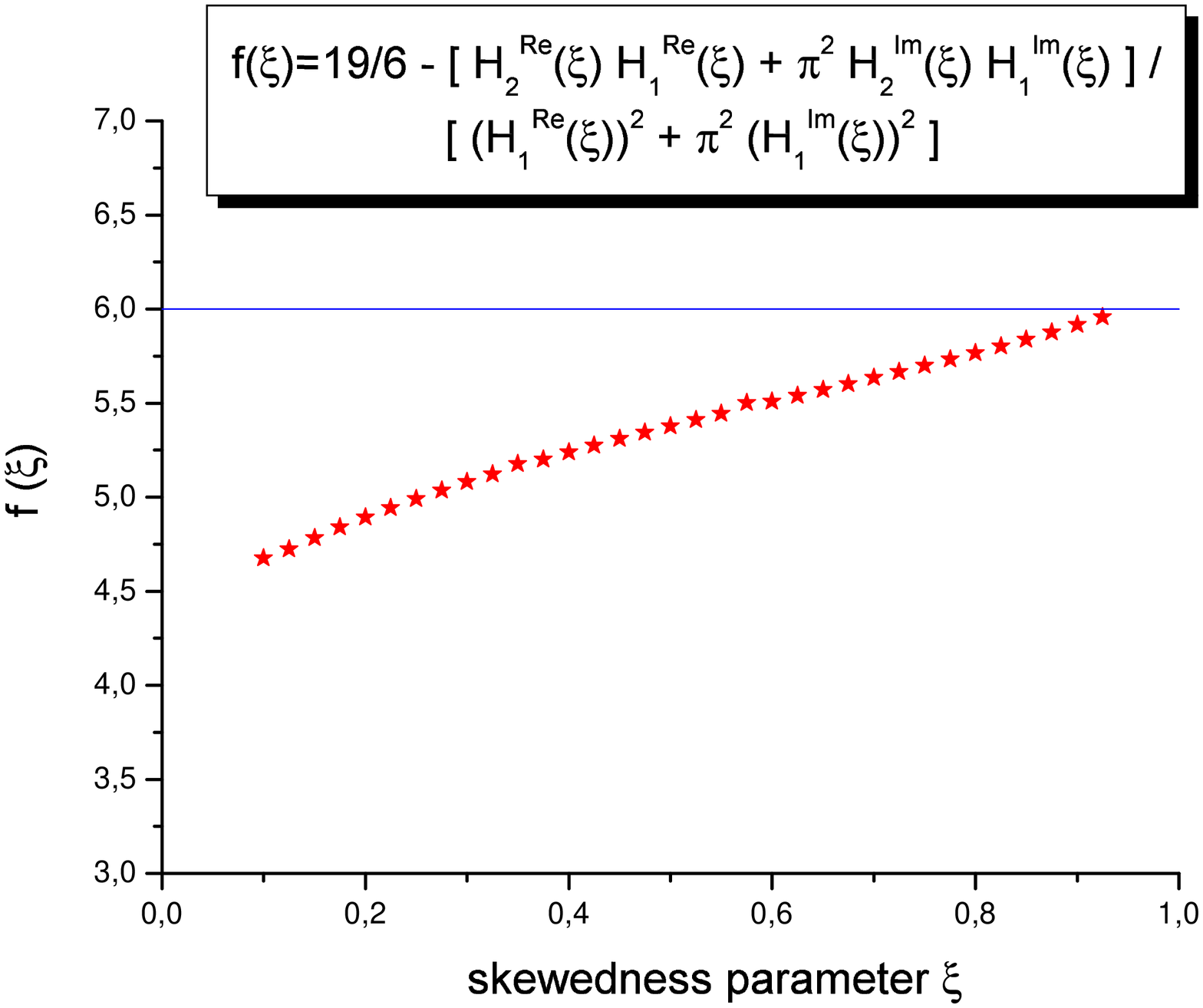}}$$
\caption{ The $\rho$ meson production (quark contribution only): the BLM scales extracted
from the squared amplitude.}
\label{blmmo1}
\end{figure}
\begin{figure}
$$\rotatebox{270}{\includegraphics[width=8cm]{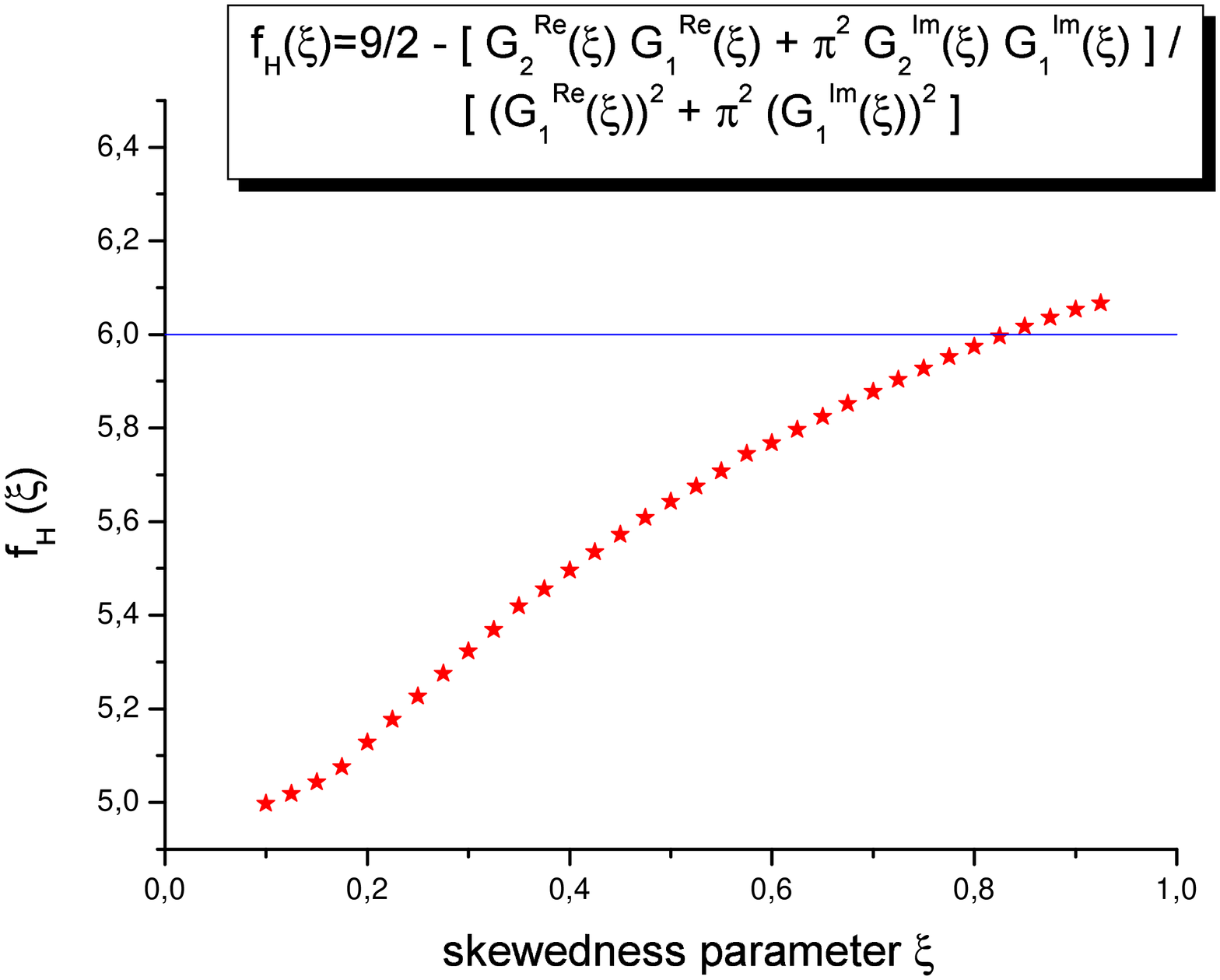}}$$
\caption{The hybrid meson production: the BLM scales extracted
from the squared amplitude.}
\label{blmmo2}
\end{figure}

\noindent
In this section, we will now extract the corresponding BLM scales working with the squared 
amplitudes or, in other words, with the cross section. In this case, the BLM equation
will be rewritten in the following form:
\begin{eqnarray}
\label{blm-sq}
\alpha_S^3(\mu^2_0)\,\frac{\beta_0}{4\pi}\,{\rm ln}\frac{\mu^2}{\mu^2_0}
\,|{\cal A}^{{\rm LO}}|^2 -
\alpha_S^3(\mu^2_0)\, \frac{\beta_0}{4\pi} \Biggl\{
 \Re e {\cal A}^{{\rm LO}}\, \Re e{\cal A}^{{\rm NLO},(\beta)} +
\Im m{\cal A}^{{\rm LO}}\, \Im m{\cal A}^{{\rm NLO},(\beta)}\Biggr\}=0,
\nonumber\\
\end{eqnarray}
where we introduced the notations
\begin{eqnarray}
{\cal A}^{{\rm NLO},(\beta)}=
\biggl[ C - {\rm ln}\frac{Q^2}{\mu^2} \biggr] \,{\cal A}^{{\rm LO}} +
\tilde{\cal A}^{{\rm NLO},(\beta)}.
\end{eqnarray}   
From (\ref{blm-sq}), we can obtain for the $\rho$ meson function $f^{\rho}(\xi)$ and for
the hybrid meson function $f^{H}(\xi)$ the following expressions:
\begin{eqnarray}
\label{fmo}
&&f^\rho (\xi)=\frac{19}{6}-
\frac{H^{Re}_2(\xi)H^{Re}_1(\xi)+\pi^2H^{Im}_2(\xi)H^{Im}_1(\xi)}
{(H^{Re}_1(\xi))^2+\pi^2 (H^{Im}_1(\xi))^2}
\end{eqnarray}
and
\begin{eqnarray}
\label{fmoH}
&&f^H(\xi)=\frac{9}{2}-
\frac{G^{Re}_2(\xi)G^{Re}_1(\xi)+\pi^2G^{Im}_2(\xi)G^{Im}_1(\xi)}
{(G^{Re}_1(\xi))^2+\pi^2 (G^{Im}_1(\xi))^2}.
\end{eqnarray}
In (\ref{fmo}) and (\ref{fmoH}), the structure functions 
$H^{Re}_i(\xi)$, $H^{Im}_i(\xi)$, $G^{Re}_i(\xi)$ and $G^{Im}_i(\xi)$ are the same as they
were defined for the BLM scales (\ref{fre}) -- (\ref{fimH}).  
The curves of (\ref{fmo}) and (\ref{fmoH}) are shown on Fig. \ref{blmmo1}
and \ref{blmmo2}.

\vspace{.5cm}
\section*{Conclusions}
\vspace{.5cm}

\noindent
We have shown that the usual way of applying the BLM method to the scattering amplitude
of exclusive vector meson electroproduction  
leads to equations fixing the BLM scale which are singular; 
we believe that this invalidates their 
straightforward use. The reason that  this problem has not been taken 
care of in previous studies \cite{BLPS} comes from the fact that these 
studies mostly concentrated on the simpler case of the meson form 
factor and of amplitudes which were rewritten in terms of this 
quantity. Let us stress that the singular behaviour of the equations is 
not specific to QCD. Indeed a simple calculation shows that the same 
behaviour would be obtained in a QED calculation where gluons are 
replaced by photons, quarks by electrons, and the $\rho$ meson  by a 
positronium bound state described by a conveniently defined 
distribution amplitude. We know that it is often advocated 
that the BLM procedure is easier to understand in the abelian case than 
in the non-abelian one, but its application to an exclusive process 
where the amplitude contains both a real and imaginary parts 
nevertheless  suffers from the problem outlined in this paper. 
We have demonstrated that such singularities do not appear if the BLM procedure is applied 
to the square of the scattering amplitude, which is a quantity more closely 
related to an observable, the cross section, rather than the scattering amplitude itself.
The phenomenological consequences for the vector meson electroproduction 
based on this new way of fixing of the BLM scale is studied in \cite{APSTW2}.

\vspace{.5cm}
\section*{Acknowledgments}
\vspace{.5cm}

\noindent 
We acknowledge useful discussions with A.~Bakulev, G.~Grunberg, A.~Kataev,
G.~Korchemsky, and, especially, S.~Mikhailov.
This work is supported in part by INTAS (Project 00/587), RFBR (Grant 03-02-16816)
and by the Polish Grant  1 P03B 028 28.
The work of B.~P., L.~Sz. and S.~W. is partially supported by the French-Polish scientific 
agreement Polonium and the Joint Research Activity "Generalized Parton 
Distributions" of the european I3 program Hadronic Physics, contract 
RII3-CT-2004-506078. 
I.~V.~A. thanks  NATO for a Grant.
L.~Sz. is a Visiting Fellow of the Fonds National pour la Recherche Scientifique (Belgium).

\vspace{.5cm}
\section*{Appendix: Typical functions for the
determination of BLM scales}
\vspace{.5cm}

\noindent
The typical functions in terms of which the renormalization scales corresponding to
the $\rho$ and hybrid mesons are rewritten  read
\begin{eqnarray}
\label{Hre2}
H^{Re}_2(\xi)&=&{\cal P}\int\limits_{-1}^{1}dx H^{p}_{\rho^0}(x,\xi,t_{min})
\Biggl[
{\rm ln}\biggl|\frac{\xi+x}{2\xi}\biggr|\frac{1}{\xi+x}-
{\rm ln}\biggl|\frac{\xi-x}{2\xi}\biggr|\frac{1}{\xi-x}
\Biggr] +
\nonumber\\
&&\frac{\pi^2}{2}
\biggl[H^{p}_{\rho^0}(-\xi,\xi,t_{min})-H^{p}_{\rho^0}(\xi,\xi,t_{min})\biggr];
\\
\label{Hre1}
H^{Re}_1(\xi)&=&{\cal P}\int\limits_{-1}^{1}dx H^{p}_{\rho^0}(x,\xi,t_{min})
\Biggl[\frac{1}{\xi+x}-\frac{1}{\xi-x}
\Biggr],
\end{eqnarray}
and
\begin{eqnarray}
\label{Him2}
H^{Im}_2(\xi)&=&\int\limits_{\xi}^{1}dx
\frac{H^{p}_{\rho^0}(x,\xi,t_{min})-H^{p}_{\rho^0}(\xi,\xi,t_{min})}{\xi-x}-
\nonumber\\
&&\int\limits_{-1}^{-\xi}dx
\frac{H^{p}_{\rho^0}(x,\xi,t_{min})-H^{p}_{\rho^0}(-\xi,\xi,t_{min})}{\xi+x}
\nonumber\\
&&+{\rm ln}\biggl|\frac{1-\xi}{2\xi}\biggr|
\biggl[H^{p}_{\rho^0}(-\xi,\xi,t_{min})-H^{p}_{\rho^0}(\xi,\xi,t_{min})\biggr];
\\
\label{Him1}
H^{Im}_1(\xi)&=&\biggl[H^{p}_{\rho^0}(-\xi,\xi,t_{min})-H^{p}_{\rho^0}(\xi,\xi,t_{min})\biggr]
\end{eqnarray}
for the $\rho$ meson production, and
\begin{eqnarray}
\label{Hre2H}
G^{Re}_2(\xi)&=&{\cal P}\int\limits_{-1}^{1}dx H^{p}_H(x,\xi,t_{min})
\Biggl[
{\rm ln}\biggl|\frac{\xi+x}{2\xi}\biggr|\frac{1}{\xi+x}+
{\rm ln}\biggl|\frac{\xi-x}{2\xi}\biggr|\frac{1}{\xi-x}
\Biggr] +
\nonumber\\
&&\frac{\pi^2}{2}
\biggl[H^{p}_H(-\xi,\xi,t_{min})+H^{p}_H(\xi,\xi,t_{min})\biggr];
\\
\label{Hre1H}
G^{Re}_1(\xi)&=&{\cal P}\int\limits_{-1}^{1}dx H^{p}_H(x,\xi,t_{min})
\Biggl[\frac{1}{\xi+x}+\frac{1}{\xi-x}
\Biggr],
\end{eqnarray}
and
\begin{eqnarray}
\label{Him2H}
G^{Im}_2(\xi)&=&-\int\limits_{\xi}^{1}dx
\frac{H^{p}_H(x,\xi,t_{min})-H^{p}_H(\xi,\xi,t_{min})}{\xi-x}-
\nonumber\\
&&\int\limits_{-1}^{-\xi}dx
\frac{H^{p}_H(x,\xi,t_{min})-H^{p}_H(-\xi,\xi,t_{min})}{\xi+x}
\nonumber\\
&&+{\rm ln}\biggl|\frac{1-\xi}{2\xi}\biggr|
\biggl[H^{p}_H(-\xi,\xi,t_{min})+H^{p}_H(\xi,\xi,t_{min})\biggr];
\\
\label{Him1H}
G^{Im}_1(\xi)&=&\biggl[H^{p}_H(-\xi,\xi,t_{min})+H^{p}_H(\xi,\xi,t_{min})\biggr]
\end{eqnarray}
for the hybrid meson production.

%%%%%%%%%%%%%%%%%%%%%%%%%%%%%%%%%%%%%%%%%%%%%%%%%%%%%%%%%%%%%%%%%%%%%%%%%%%%


\begin{thebibliography}{99}

\bibitem{opt}
P.~M.~Stevenson,
%``Optimized Perturbation Theory,''
Phys.\ Rev.\ D {\bf 23} (1981) 2916;
%%CITATION = PHRVA,D23,2916;%%
P.~M.~Stevenson,
%``Resolution Of The Renormalization Scheme Ambiguity In Perturbative QCD,''
Phys.\ Lett.\ B {\bf 100}  (1981) 61;
%%CITATION = PHLTA,B100,61;%%
G.~Grunberg,
%``Renormalization Scheme Independent QCD And QED: The Method Of Effective
%Charges,''
Phys.\ Rev.\ D {\bf 29} (1984) 2315.
%%CITATION = PHRVA,D29,2315;%%

\bibitem{BLM}
S.~J.~Brodsky, G.~P.~Lepage and P.~B.~Mackenzie,
%``On The Elimination Of Scale Ambiguities In Perturbative Quantum
%Chromodynamics,''
Phys.\ Rev.\ D {\bf 28} 228 (1983) 228.
%%CITATION = PHRVA,D28,228;%%

\bibitem{Belitsky:2001nq}
A.~V.~Belitsky and D.~Muller,
%``Hard exclusive meson production at next-to-leading order,''
Phys.\ Lett.\ B {\bf 513} (2001) 349.
%%CITATION = HEP-PH 0105046;%%

\bibitem{IS}
D.~Y.~Ivanov, L.~Szymanowski and G.~Krasnikov,
%``Vector meson electroproduction at next-to-leading order,''
JETP Lett.\  {\bf 80} (2004) 226
[Pisma Zh.\ Eksp.\ Teor.\ Fiz.\  {\bf 80} (2004) 255]
[arXiv:hep-ph/0407207].
%%CITATION = HEP-PH 0407207;%%


\bibitem{ffNLO}
A.~V.~Radyushkin,
%``Asymptotic Behavior Of The Pion Form-Factor In QCD. (In Russian),''
Fiz.\ Elem.\ Chast.\ Atom.\ Yadra {\bf 20} (1989) 97;
%%CITATION = FECAA,20,97;%%
R.~D.~Field, R.~Gupta, S.~Otto and L.~Chang,
%``Beyond Leading Order QCD Perturbative Corrections To The Pion Form-Factor,''
Nucl.\ Phys.\ B {\bf 186} (1981) 429;
%%CITATION = NUPHA,B186,429;%%
F.~M.~Dittes and A.~V.~Radyushkin,
%``Radiative Corrections To The Pion Form-Factor In Quantum Chromodynamics. (In
%Russian),''
Sov.\ J.\ Nucl.\ Phys.\  {\bf 34} (1981) 293 [Yad.\ Fiz.\  {\bf 34} (1981) 529].
%%CITATION = SJNCA,34,293;%%

\bibitem{CFS}
J.~C.~Collins, L.~Frankfurt and M.~Strikman,
%``Factorization for hard exclusive electroproduction of mesons in QCD,''
Phys.\ Rev.\ D {\bf 56} (1997)  2982
[arXiv:hep-ph/9611433].
%%CITATION = HEP-PH 9611433;%%

\bibitem{RadE}
A.~V.~Radyushkin,
%``Asymptotic Behavior Of The Pion Form-Factor In QCD. (In Russian),''
Fiz.\ Elem.\ Chast.\ Atom.\ Yadra {\bf 20} (1989) 97.
%%CITATION = FECAA,20,97;%%

\bibitem{APSTW1}
I.~V.~Anikin, B.~Pire, L.~Szymanowski, O.~V.~Teryaev and S.~Wallon,
%``Deep electroproduction of exotic hybrid mesons,''
Phys.\ Rev.\ D {\bf 70} (2004) 011501
%[arXiv:hep-ph/0401130].
%%CITATION = HEP-PH 0401130;%%

\bibitem{Rad}
A.~V.~Radyushkin,
%``Double distributions and evolution equations,''
Phys.\ Rev.\ D {\bf 59} (1999) 014030.
%%CITATION = HEP-PH 9805342;%%

\bibitem{MRST98}
A.~D.~Martin, R.~G.~Roberts, W.~J.~Stirling and R.~S.~Thorne,
%``Parton distributions: A new global analysis,''
Eur.\ Phys.\ J.\ C {\bf 4} (1998) 463
[arXiv:hep-ph/9803445].
%%CITATION = HEP-PH 9803445;%%

\bibitem{APSTW2}
I.~V.~Anikin, B.~Pire, L.~Szymanowski, O.~V.~Teryaev and S.~Wallon,
%``Exotic hybrid mesons in hard electroproduction,''
Phys.\ Rev.\ D {\bf 71} (2005) 034021
[arXiv:hep-ph/0411407].
%%CITATION = HEP-PH 0411407;%%

\bibitem{PW-LD}
M.~V.~Polyakov and C.~Weiss,
%``Skewed and double distributions in pion and nucleon,''
Phys.\ Rev.\ D {\bf 60}  (1999) 114017;
%%CITATION = HEP-PH 9902451;%%
B.~Lehmann-Dronke, A.~Sch{\"a}fer, M.~V.~Polyakov and K.~Goeke,
%``Angular distributions in hard exclusive production of pion pairs,''
Phys.\ Rev.\ D {\bf 63} (2001) 114001.
%%CITATION = HEP-PH 0012108;%%

\bibitem{BLPS}
S.~J.~Brodsky, C.~R.~Ji, A.~Pang and D.~G.~Robertson,
%``Optimal renormalization scale and scheme for exclusive processes,''
Phys.\ Rev.\ D {\bf 57} (1998) 245
[arXiv:hep-ph/9705221].
%%CITATION = HEP-PH 9705221;%%


\end{thebibliography}
\end{document}